\RequirePackage{lineno}
\documentclass[aps,twocolumn,showpacs,byrevtex,prl,reprint,nofootinbib]{revtex4-1}
\usepackage{xspace}
\usepackage[english]{babel}

\usepackage{graphicx}
\usepackage{dcolumn}
\usepackage{bm}
\usepackage{rotating}
\usepackage{color}
\usepackage{verbatim} 
\usepackage{multirow}
\usepackage{amsmath}

\begin{document}

\title{\boldmath  Study of $\eta(1475)$ and $X(1835)$  in radiative $J/\psi$ decays to $\gamma \phi$}

\author{
  \small
M.~Ablikim$^{1}$, M.~N.~Achasov$^{9,e}$, S. ~Ahmed$^{14}$, O.~Albayrak$^{5}$, M.~Albrecht$^{4}$, M.~Alekseev$^{51A,51C}$, D.~J.~Ambrose$^{46}$, A.~Amoroso$^{51A,51C}$, F.~F.~An$^{1}$, Q.~An$^{48,a}$
% ...
      \\
      \vspace{0.2cm}
      (BESIII Collaboration)\\
      \vspace{0.2cm} {\it
      $^{1}$ Institute of High Energy Physics, Beijing 100049, People's Republic of China\\
% ...
$^{4}$ Bochum Ruhr-University, D-44780 Bochum, Germany\\
$^{5}$ Carnegie Mellon University, Pittsburgh, Pennsylvania 15213, USA\\
% ...
$^{9}$ G.I. Budker Institute of Nuclear Physics SB RAS (BINP), Novosibirsk 630090, Russia\\
% ...
$^{14}$ Helmholtz Institute Mainz, Johann-Joachim-Becher-Weg 45, D-55099 Mainz, Germany\\
% ...
$^{46}$ University of Rochester, Rochester, New York 14627, USA\\
% ...
$^{48}$ University of Science and Technology of China, Hefei 230026, People's Republic of China\\
% ...
$^{51}$ (A)University of Turin, I-10125, Turin, Italy; (B)University of Eastern Piedmont, I-15121, Alessandria, Italy; (C)INFN, I-10125, Turin, Italy\\
% ...
$^{a}$ Also at State Key Laboratory of Particle Detection and Electronics, Beijing 100049, Hefei 230026, People's Republic of China\\
% ...
$^{e}$ Also at the Novosibirsk State University, Novosibirsk, 630090, Russia\\
% ...
      }
}

\begin{abstract}

       The decay $J/\psi \rightarrow \gamma \gamma \phi$ is studied using a sample of $1.31\times10^{9}$ $J/\psi$ events collected with the BESIII detector. Two structures around 1475 MeV/c$^2$ and 1835 MeV/c$^2$ are observed in the $\gamma \phi$ invariant mass spectrum for the first time. With a fit on the $\gamma \phi$ invariant mass, which takes into account the interference between the two structures, and a simple analysis of the angular distribution, the structure around 1475 MeV/c$^2$ is found to favor an assignment as the $\eta(1475)$ and the mass and width for the structure around 1835 MeV/c$^2$ are consistent with the $X(1835)$. The statistical significances of the two structures are $13.5 \sigma$ and $6.3 \sigma$, respectively. The results indicate that both $\eta(1475)$ and $X(1835)$ contain a sizeable $s\bar{s}$ component.

\end{abstract}


\maketitle

A puzzling state, the $\eta(1440)$, was first observed in $p \bar p$ annihilation at rest into $\eta(1440)\pi^+ \pi^- \,(\eta(1440) \rightarrow K \bar K \pi$) ~\cite{eta1440obs}, and later in $J/\psi$ radiative decays to $K \bar K \pi$~\cite{eta1475obs}, $\gamma \rho$~\cite{eta1405obs} and $f_0(980) \pi^0$~\cite{eta1475wuz}. Further studies by different experiments reported evidence for the existence of two pseudo-scalar mesons in this region, the $\eta(1405)$ and the $\eta(1475)$~\cite{pdg}.
After about 50 years since the first observation of $\eta(1440)$, its structure is still an open question.
According to theoretical predictions, the $\eta(1475)$ could be interpreted as the first radial excitation of the $\eta^{\prime}$ while the $\eta(1405)$ is an excellent candidate for a $0^{-+}$ glueball in the fluxtube model~\cite{fluxtubemodel} (though this assignment of the $\eta(1405)$ is not favored by lattice gauge theories, which predict that the $0^{-+}$ glueball should be above $2$ GeV/$c^2$~\cite{lattice1,lattice2}).
However, the existence of two pseudo-scalar mesons in this region remains controversial.
 The spectrum could consist of a single state, the $\eta(1440)$, that splits due to nodes in the decay amplitudes, with the $\eta(1440)$ being the SU(3) flavor partner of the $\eta(1295)$~\cite{onestat1, onestat2, onestat3}.
Under the one-state assumption, the partial width relationship between its $\gamma \rho$ and $\gamma \phi$ decay modes is predicted to be $\Gamma_{\gamma\rho}:\Gamma_{\gamma\phi} \simeq 3.8:1$~\cite{onestat2}.

The $X(1835)$ was first observed by the BESII experiment in the $\pi\pi\eta^{\prime}$~\cite{X1835II} invariant mass spectrum and was recently confirmed with higher statistical significance by the BESIII collaboration~\cite{X1835I}.
It was also observed in the $K_S^0 K_S^0 \eta$ invariant mass spectrum by BESIII~\cite{X1835III}.
Furthermore, a recent BESIII result observes an anomalous line shape of the $X(1835)$ near the $p \bar{p}$ threshold in the decay $J/\psi \rightarrow \gamma \pi^+ \pi^- \eta^{\prime}$~\cite{baihe}.
The Belle collaboration reported an upper limit on the product $\Gamma_{\gamma \gamma}\,B(X \to \pi^+ \pi^- \eta')$ for the $X(1835)$ at the $90\%$ confidence level as 35.6\,(83) eV/$c^2$, assuming constructive (destructive) interference between the $X(1835)$ and the $\eta(1475)$~\cite{X1835IV}.
As a state with $J^{PC} = 0^{-+}$, the nature of the $X(1835)$ is still an open question, though a number of theoretical interpretations have been proposed, including an $N \bar N$ bound state~\cite{NNbstate}, baryonium with a sizable gluon content~\cite{BGLbstate}, a pseudo-scalar glueball~\cite{Glueballbstate}, a radial excitation of the $\eta'$~\cite{exetap}, and an $\eta_c$-glueball mixture~\cite{Glueballbstate}.
So far, none of these interpretations have been ruled out or confirmed.

Since radiative decays like $J/\psi\rightarrow\gamma X$, where $X \rightarrow \gamma V$ with $V = \rho$ or $\phi$, do not change the flavor structure of the intermediate states,
the final-state vector mesons $V$ act as a flavor filter, helping to understand the flavor contents of the intermediate states $X$~\cite{filter1}.
In this paper, we present an analysis of the decay $J/\psi \rightarrow \gamma \gamma \phi$, where the $\phi$ meson is reconstructed in the $K^+K^-$ final state, based on a sample of $1.31\times10^9$ $J/\psi$ events~\cite{njpsi} collected with the BESIII detector~\cite{bibbes3}.

The BESIII detector is a magnetic spectrometer operating at the double-ring
$e^{+}e^{-}$ collider BEPCII with center-of-mass energies between 2.0 and 4.6 GeV.
The cylindrical core of the BESIII detector consists of a helium-based main drift chamber
(MDC), a plastic scintillator time-of-flight system (TOF), and a CsI(Tl) electromagnetic
calorimeter (EMC) that are all enclosed in a superconducting solenoidal magnet providing a
magnetic field of 1.0 T (0.9 T in 2012, for about $1.09 \times 10^9$ $J/\psi$ events).
The solenoid is supported by an octagonal flux-return yoke with resistive plate counter muon
identifier modules interleaved with steel.
The acceptance for charged particles and photons is 93\% of the 4${\pi}$ solid angle, and the charged-particle momentum resolution at 1 GeV/$c$ is about 0.5\%.
The EMC measures photon energies with a resolution of 2.5\% (5\%) at 1 GeV in the barrel (endcaps).
A {\footnotesize GEANT}4-based~\cite{bibgeant4} Monte Carlo (MC) simulation software package is used to optimize the event selection criteria, estimate backgrounds and determine the detection efficiencies.

Charged tracks that have a polar angle $|\cos \theta|<0.93$ and that pass within $\pm 10$ cm of the interaction point along the beam direction and within 1 cm in the plane perpendicular to the beam are accepted.
The combined information from specific energy
loss ($dE/dx$) measurements in the MDC and the flight time measured in the TOF is used to form particle identification (PID) confidence levels for the $\pi$, $K$, and $p$ hypotheses.
Each track is assigned the particle type corresponding to the highest confidence level.
Photon candidates are required to have an energy deposition above 25 MeV in the barrel EMC ($|\cos \theta|< 0.80$) or 50 MeV in the endcap EMC ($0.86 <|\cos \theta|< 0.92$).
To exclude showers from charged particles, the angle between the shower direction and the charged tracks extrapolated to the EMC must be greater than 10 degrees.
A requirement on the EMC timing ($0 \le t \le 700$ ns) is used to suppress electronic noise and energy deposits unrelated to the event of interest.

\begin{figure}[!htbp]
\centering
  \includegraphics[width=0.42\textwidth]{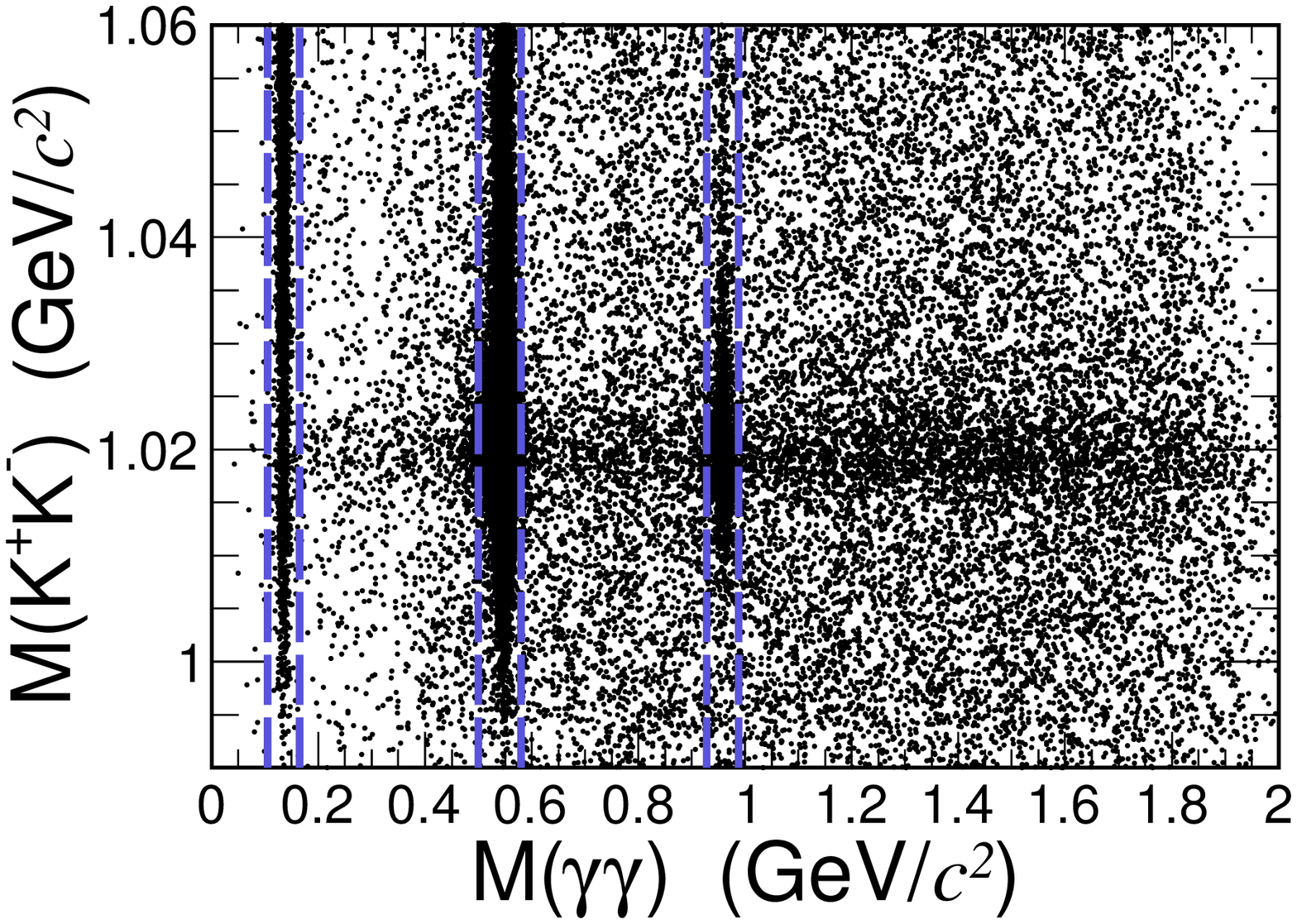}
  \put(-55,125){\textbf{\large{(a)}}}\\
  \includegraphics[width=0.42\textwidth]{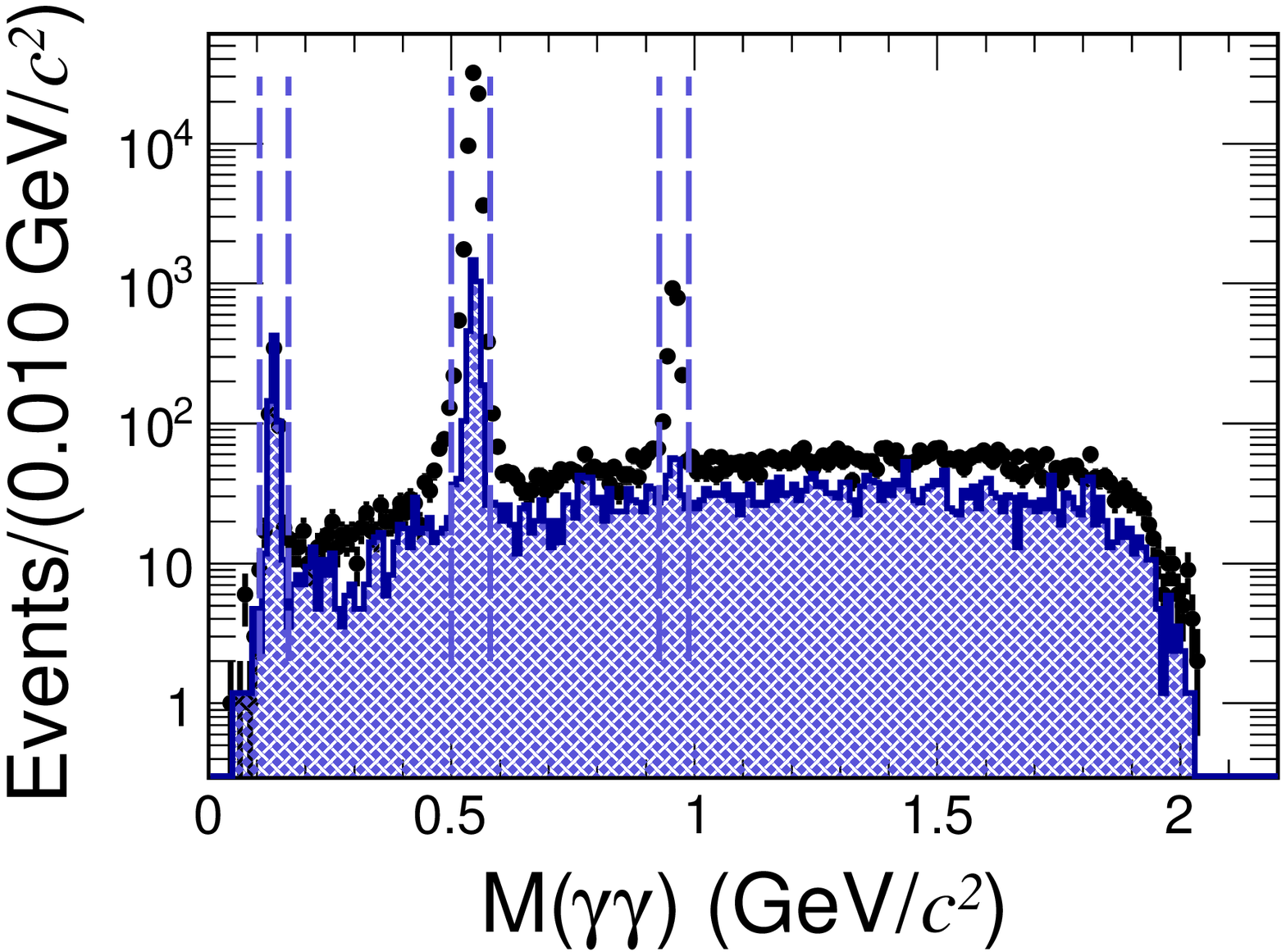}
  \put(-55,125){\textbf{\large{(b)}}} \\
  \includegraphics[width=0.42\textwidth]{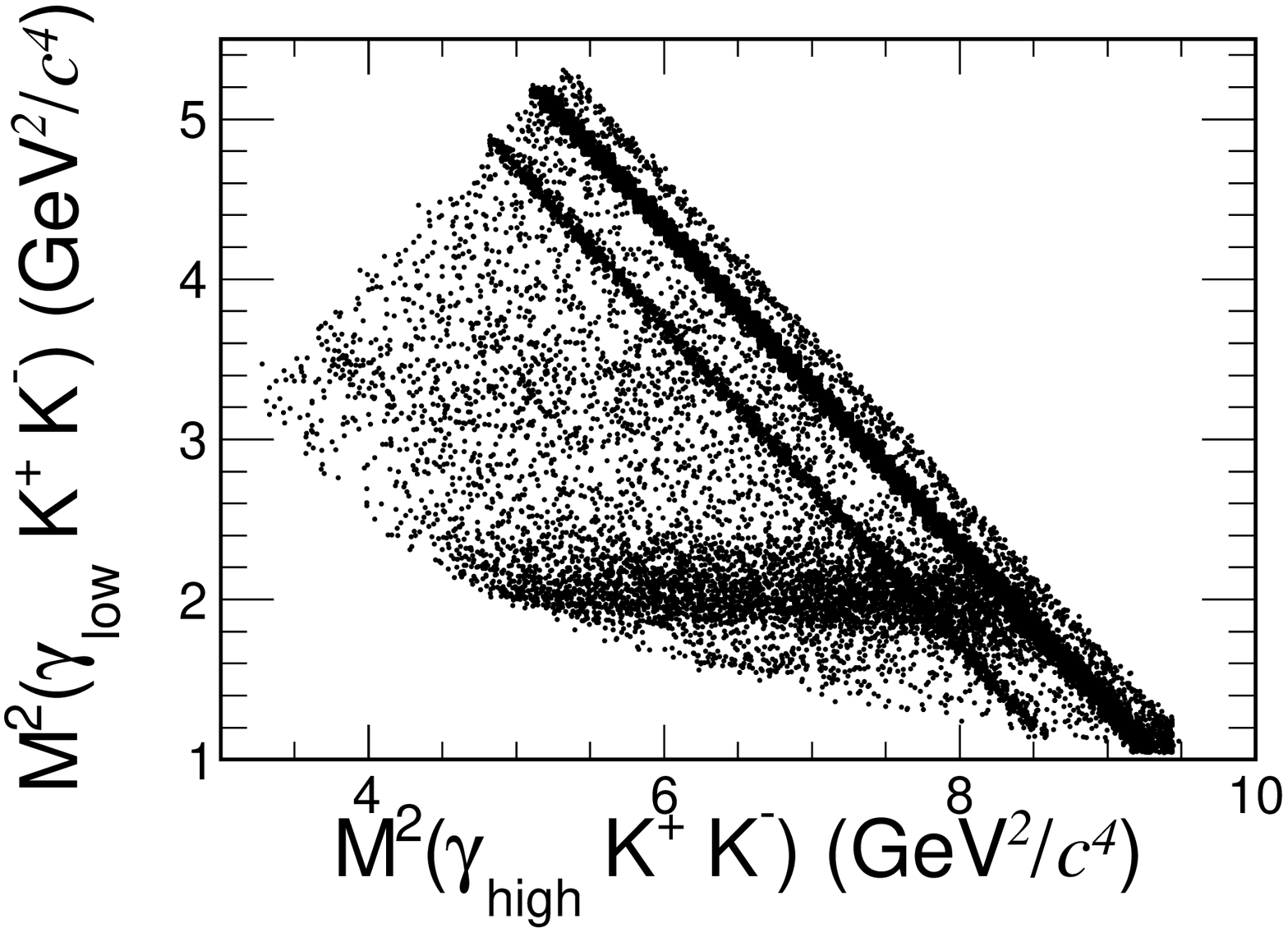}
  \put(-55,125){\textbf{\large{(c)}}}
  \caption{
  (a) Scatter plot of $M(K^+K^-)$ versus $M(\gamma\gamma)$.
              (b) Projections of $M(\gamma\gamma)$ for the events in the $\phi$ signal region (dots with error bar) and sideband regions (histogram).
              (c)  Dalitz plot of $M^2(\gamma_{\rm low}K^+K^-)$ versus $M^2(\gamma_{\rm high}K^+K^-)$.
  }
  \label{fig:scat}
\end{figure}

For the decay $J/\psi \rightarrow \gamma \gamma \phi\,(\phi \rightarrow K^+ K^-$), candidate events are required to have two oppositely charged tracks identified as kaons and at least two photons.  A kinematic fit constraining the total four-momentum to the initial $J/\psi$ four-momentum ($4C$-fit) is performed under the final state hypothesis $\gamma \gamma K^+ K^-$. In candidate events with more than two photon candidates,
the combination with the minimum chi-square from the kinematic fit $\chi^2_{4C}$ is retained. Only events with $\chi^2_{4C}<40$ are accepted. To reject possible backgrounds with three or four photons in the final state, {similar 4C} kinematic fits are performed under the background hypotheses $J/\psi \rightarrow \gamma \gamma \gamma K^+ K^-$ and $J/\psi \rightarrow \gamma \gamma \gamma \gamma K^+ K^-$. The events with a $\chi^2_{4C}$ value for the signal hypothesis larger than any of those for the background hypotheses are discarded.
After applying the above selection criteria, the distribution of the $K^+ K^-$ invariant mass $M(K^+ K^-)$ versus the $\gamma\gamma$ invariant mass $M(\gamma\gamma)$ of surviving candidate events is shown in Fig.~\ref{fig:scat}~(a).  A clear horizontal band, representing the $\phi$ from the signal decay $J/\psi\to\gamma\gamma\phi$, is observed. There are also three vertical bands representing the two-photon decays of $\pi^0$, $\eta$ and $\eta^{\prime}$, which are from the backgrounds of $J/\psi\to K^+K^-\pi^0$, $K^+K^-\eta$, and  $K^+K^-\eta^{\prime}$, respectively.
The projections of $M(\gamma\gamma)$ for the events in the $\phi$ signal region defined as $|M(K^+K^-)-m(\phi)|<0.010$~GeV/$c^2$, and in the $\phi$ sideband region defined as $0.020<|M(K^+K^-)-m(\phi)|<0.030$~GeV/$c^2$ are shown in Fig.~\ref{fig:scat}~(b), individually,  where $m(\phi)$ is the world average value for the mass of the $\phi$ meson~\cite{pdg}. The much more prominent $\eta$ and $\eta^{\prime}$ signals observed in the $\phi$ signal region come from the background processes $J/\psi\to \phi \eta$ and $\phi \eta^{\prime}$, respectively. The Dalitz plot of $M^2(\gamma_{\rm low} K^+K^-)$ versus $M^2(\gamma_{\rm high} K^+K^-)$ for the events in the $\phi$ signal region is shown in Fig.~\ref{fig:scat}~(c), where $\gamma_{\rm low}$ and $\gamma_{\rm high}$ are the photons with low and high energy, respectively. Beside the expected diagonal bands for the $\pi^0$, $\eta$, and $\eta^{\prime}$ signals, there is a horizontal band with $M(\gamma_{\rm low} K^+K^-)$ around 1.47~GeV/$c^2$ that is of particular interest.
To further suppress the backgrounds discussed above, the requirements on the $M(\gamma\gamma)$ distribution, $|M(\gamma\gamma)-m(\pi^0)|>0.03$ GeV/$c^2$, $M(\gamma\gamma) <0.50$ GeV/$c^2$ or $M(\gamma\gamma) > 0.58$ GeV/$c^2$ and $|M(\gamma\gamma)-m(\eta^{\prime})|>0.03$ GeV/$c^2$, are applied, where $m(\pi^{0})$ and $m(\eta^{\prime})$ are the nominal masses of the $\pi^{0}$ and $\eta^{\prime}$ mesons~\cite{pdg}, respectively. By applying this additional requirement, the above backgrounds are reduced to negligible levels.

\begin{figure}[!htbp]
\flushleft
\centering
  \includegraphics[width=0.42\textwidth]{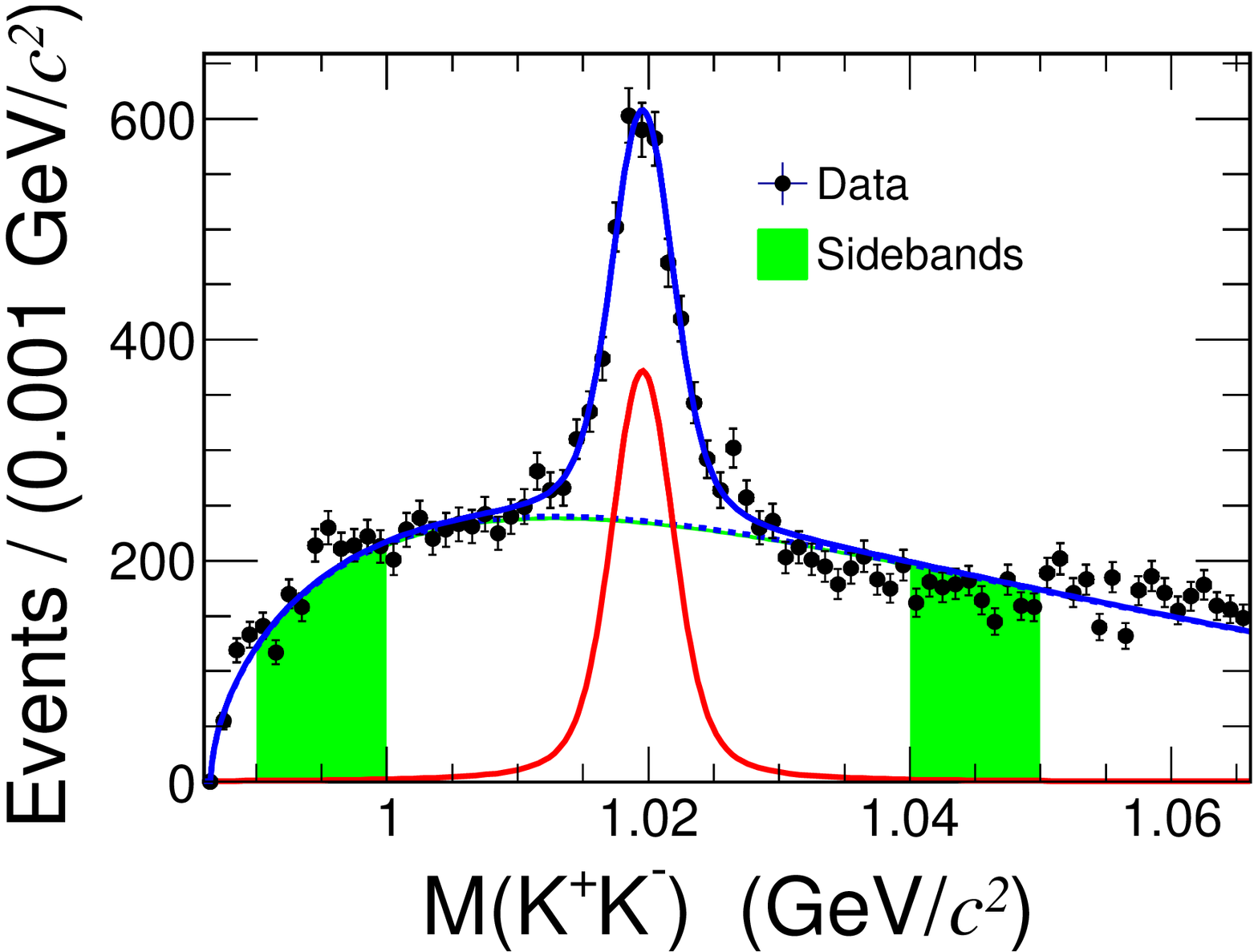}
  \put(-55,125){\textbf{\large{(a)}}}

  \includegraphics[width=0.42\textwidth]{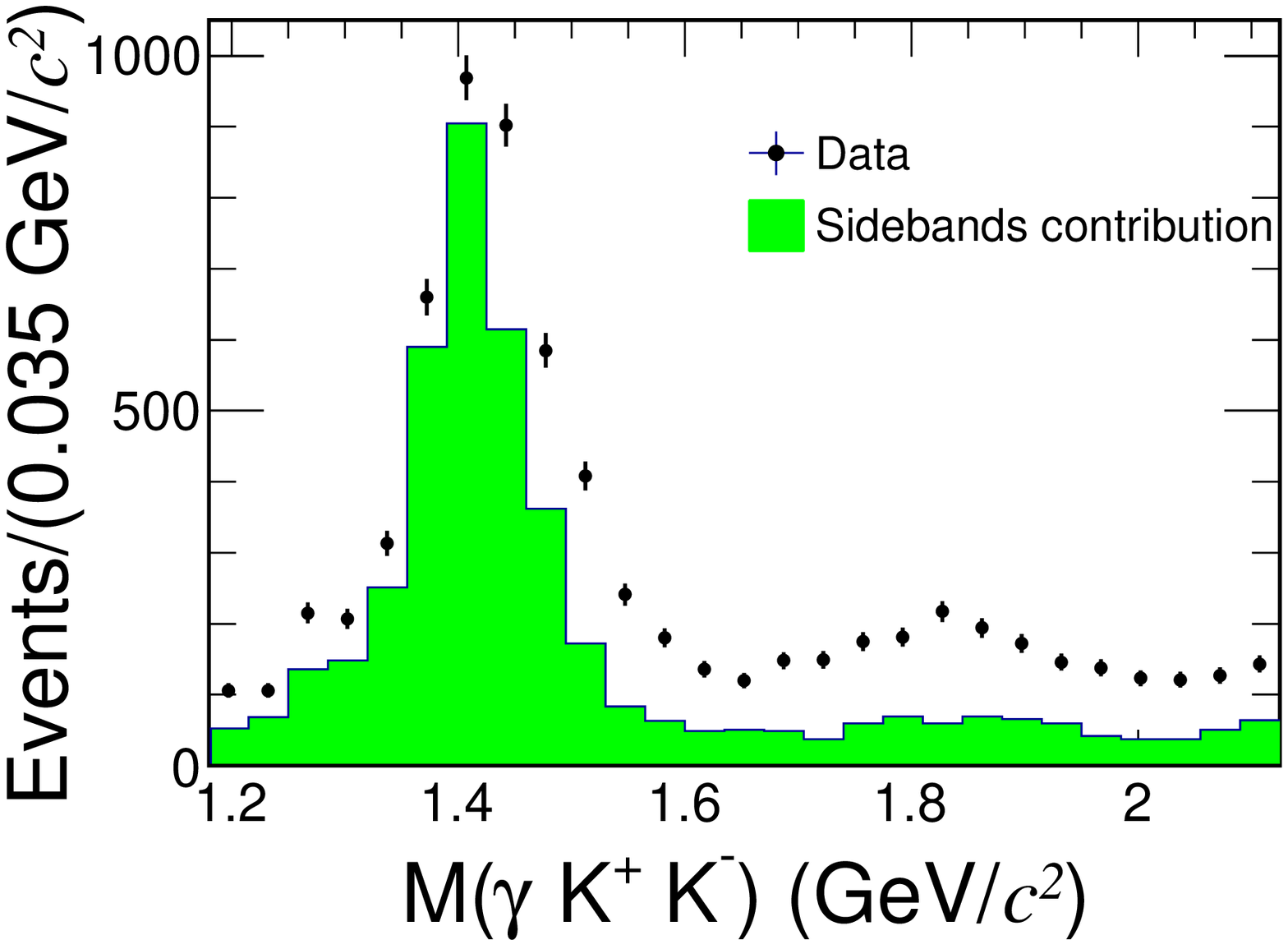}
  \put(-55,125){\textbf{\large{(b)}}}
  \caption{
  (a) Distribution of $M(K^{+}K^{-})$. The non-$\phi$ background distribution is shown with the shaded histogram.
  (b) The $M(\gamma K^+K^-)$ distribution for candidate events in the $\phi$ signal region (dots with error bars) and $\phi$ sideband region (shaded histogram).
  }
  \label{fig:scatter}
\end{figure}

After applying all of above selection criteria, the $M(K^+K^-)$ distribution is shown in Fig.~\ref{fig:scatter}~(a), in which an obvious $\phi$ signal is visible.
The distributions of the $\gamma K^+K^-$ invariant mass, $M(\gamma K^+K^-)$, two entries per event, for the event candidates in the $\phi$ signal and sideband regions are shown in Fig.~\ref{fig:scatter}~(b), where two structures near 1.47 and 1.83 GeV/$c^2$ are clearly seen in both the $\phi$ signal and sideband regions, individually.
Possible backgrounds are studied with a MC sample containing $1.2 \times 10^9$ inclusive $J/\psi$ decays, where the decays with known branching fractions are generated by {\footnotesize EVTGEN}~\cite{bibevtgen} and the remaining $J/\psi$ decays are generated according to the {\footnotesize LUNDCHARM}~\cite{biblundcharm} model.
The background without the $\phi$ meson in the final state, which is denoted as non-$\phi$ background thereafter and can be represented with the candidate events in the $\phi$ sideband region, is dominated with the decay of $J/\psi\to\gamma K^+K^- \pi^0$, where the $\pi^0$ decays asymmetrically with a low energy photon un-detected. The structure around the 1.47 GeV/$c^2$ in the $\phi$ sideband region is originated from the $J/\psi$ radiative decay to $\eta(1405)/\eta(1475)$ and $f_1(1420)$ with subsequently decaying to $\pi^0 K^+K^-$.
The background with $\phi$ meson in the final state, denoted as $\phi$ background thereafter, is dominated by the decay of $J/\psi\to\phi \pi^0\pi^0$, with two $\pi^0$ decaying asymmetrically.
The study based on a dedicated MC sample, simulated according to amplitude of $J/\psi\to\phi \pi\pi$ in Ref.~\cite{phipipi}, indicate no prominent structure appears on the $M(\gamma K^+K^-)$ distribution, though abundant structures, $e.g.$ $f_0(1500)$ and $f_0(1710)$, are on the $\pi^0\pi^0$ invariant mass distribution.

\begin{figure}[!htbp]
  \centering
  \vskip -0.0cm
  \includegraphics[width=0.42\textwidth]{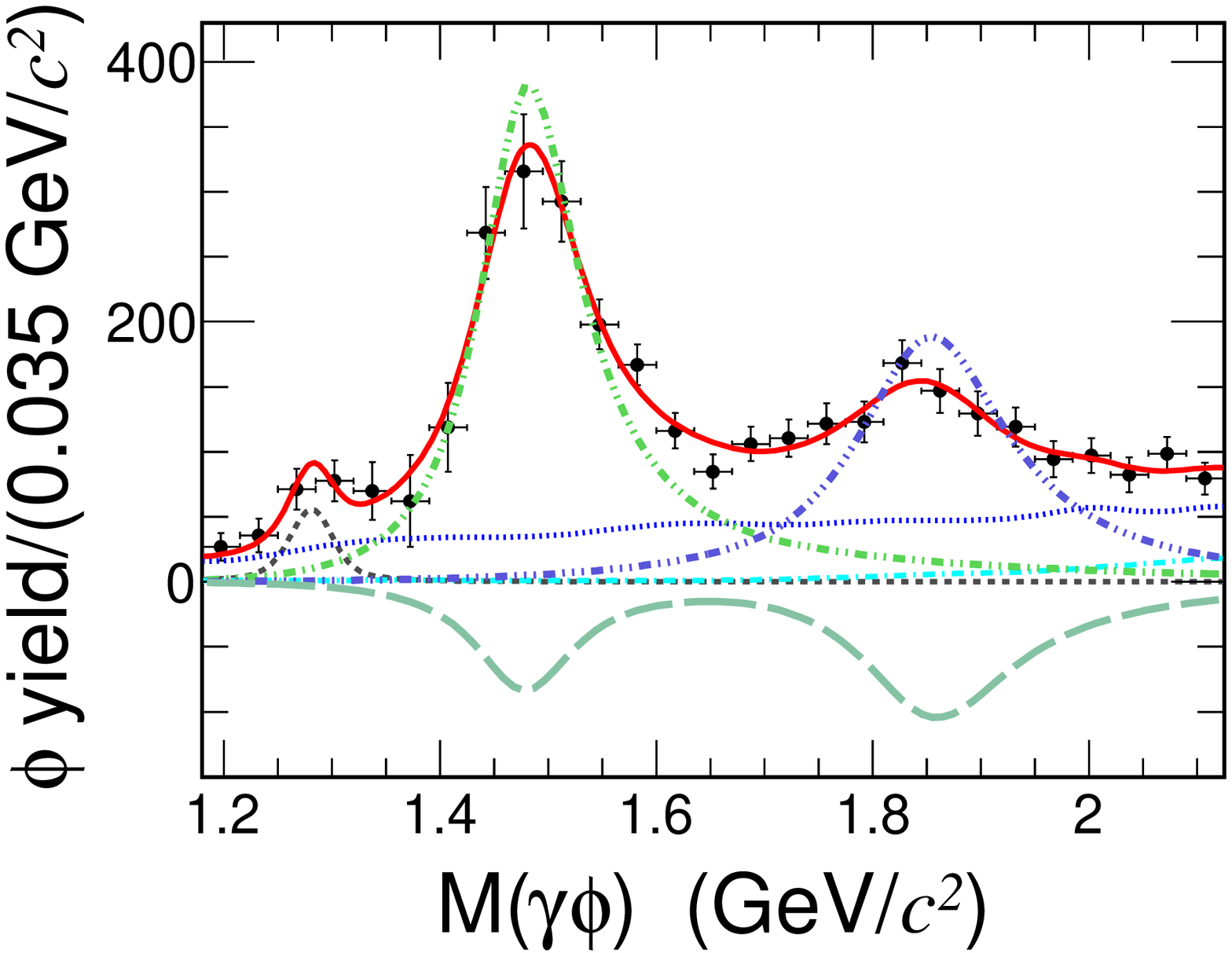}
  \put(-55,130){\textbf{\large{(a)}}}

  \includegraphics[width=0.42\textwidth]{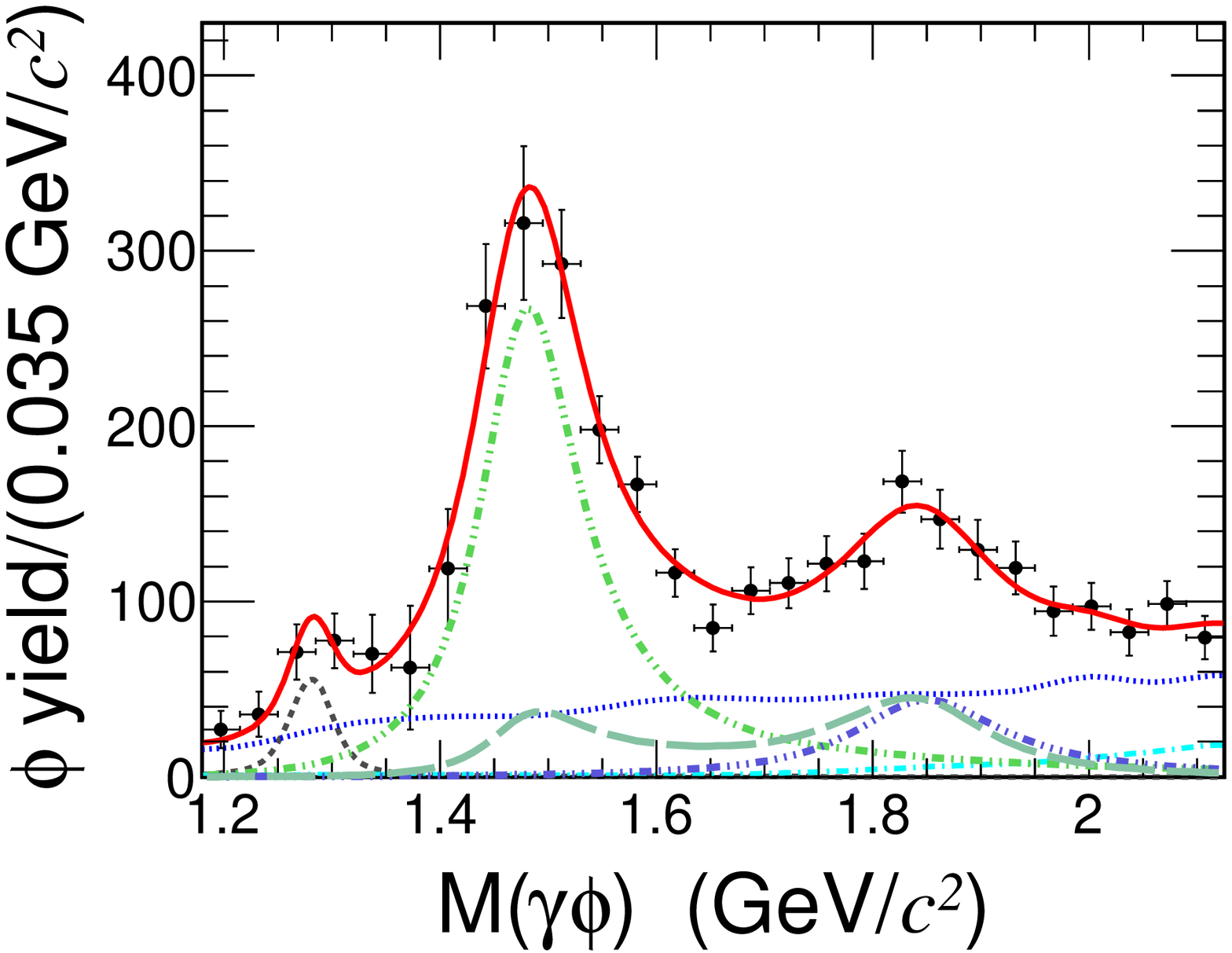}
  \put(-55,130){\textbf{\large{(b)}}}\\
  \caption{Fits to the $M_{\gamma\phi}$ distributions~(two combinations per event) for the case of (a) constructive and (b) destructive interference. The dots with error bars are the data. The (red) solid, (green) dash double-dotted, dash triple-dotted, (black) dashed, (blue) dotted and long-dashed lines are the fit results, the structures around 1.47, 1.83 GeV/$c^2$, $f_1(1285)$, backgrounds and interference contributions, respectively.}
  \label{fig:fit}
\end{figure}

To determine the signal yields for $J/\psi\to\gamma\gamma\phi$, we perform maximum-likelihood fits to the $M(K^+K^-)$ distribution in bins of $M(\gamma K^+K^-)$, called in the following the `bin-by-bin fit', where two combinations of $\gamma K^+K^-$ are considered per event.
In the fit, the $\phi$ signal is described by the MC simulated shape convolved with a Gaussian function to take into account the difference of the resolutions between the data and MC simulation.
An ARGUS function~\cite{bibargus} is used to model the non-$\phi$ backgrounds.
Interference effects between the non-$\phi$ background and the $\phi$ signal are not considered.
The signal yields as a function of the $\gamma\phi$ invariant mass $M(\gamma\phi)$ are shown in Fig.~\ref{fig:fit}.
Except for the two prominent structures around 1.47 and 1.83~GeV/$c^2$, there is small bump around 1.3~GeV/c$^2$, which is assumed to be the $f_{1}(1285)$ due to the small statistics.

A binned least-$\chi^2$ fit to the obtained $M(\gamma \phi)$ distribution is performed, in which the contribution of three resonant structures and the background from $J/\psi\to\phi \pi^0\pi^0$ are included. The direct double radiative decay $J/\psi\to\gamma\gamma \phi$ is expected to be very small, and is expected from MC studies to show a similar $M(\gamma \phi)$ distribution as that of background $J/\psi\to\phi \pi^0\pi^0$; these two background contributions cannot be distinguished. Thus, the direct double radiative decay $J/\psi\to\gamma\gamma \phi$ is not explicitly considered.
In the fit, the resonant structure
is described by a Breit-Wigner function,
\begin{equation}\label{BW}
  BW_R(s) = \frac{1}{m_{R}^2-s-i\Gamma_{R} m_{R}},
\end{equation}
where $s$ denotes the square of $M(\gamma \phi)$.
The amplitudes for the $f_1(1285)$ and the two structures around 1.47 and 1.83 GeV/$c^2$,
are denoted as $BW_0$, $BW_1$, and $BW_2$ thereafter, respectively.
The overall probability density function (PDF) for the three resonant structures incorporating the effects of mass resolution $G(m_0, \sigma(s))$ and detection efficiency $\varepsilon(s)$ obtained by the MC simulation is

\begin{equation}
\begin{split}
\label{INBW}
  BW_{\rm total} = (&BW^2_0(s) + |A_1 \times BW_1(s) \\
   &+ A_2 \times BW_2(s) \times e^{i\varphi}|^2) \\
   & \otimes G(m_0, \sigma(s))\times\varepsilon(s),
\end{split}
\end{equation}
where the interference between $BW_1$ and $BW_2$ with a relative phase $\varphi$ is taken into account, and the interference between $BW_0$ and $BW_1$ ($BW_2$) is not considered due to the low statistics of $f_1(1285)$.
In Eq.~\ref{INBW}, $A_1$ and $A_2$ are the corresponding strengths relative to $f_1(1285)$, and are determined in the fit.
In the fit, the mass and width of $f_1(1285)$ are fixed to the world average values~\cite{pdg},
while the masses and widths of $BW_1$ and $BW_2$ are free parameters.
The shape of the background $J/\psi\to\phi \pi^0\pi^0$ is modeled using the distribution obtained from a dedicated MC sample.
Since two entries of $M(\gamma K^+K^-)$ per event are implemented  in the $\phi$ signal extraction, a fraction of events have the invariant mass of $\phi$ and $\gamma$ originated from the $J/\psi$ radiative decays within the fit range of the $M(\gamma\phi)$ spectrum. Thus in the fit on the $M(\gamma\phi)$ distribution, a corresponding term is also included in the fit by taking the shapes from the signal MC simulation and constraining the amplitude according to the yields of three resonances.

 Under different assumptions for the interference, two solutions with equal fit quality are found in the fit.
 The resultant fit curves are shown in Figs.~\ref{fig:fit} (a) and (b), respectively.
  The statistical significance of each resonance is determined by the changes of $\chi^2$ and degrees of freedom (d.o.f) obtained from the fits with and without the corresponding amplitude of interest included; they are found to be $13.5\sigma$ and $6.3\sigma$ for the structures around 1.47 and 1.83 GeV/$c^2$, respectively.
 The relative phase between the two structures is $273.3^\circ \pm 37.8^\circ$ for the case of constructive interference (solution $\textrm{I}$) and $118.6^\circ \pm 12.0^\circ$ for the case of destructive interference (solution $\textrm{II}$).
 The signal yields for the $f_1(1285)$ and the other two resonances around 1.47 and 1.83 GeV/$c^2$ are determined to be  $97 \pm 31$, $1327 \pm 173$ and $305 \pm 61$ for solution I, and $97 \pm 31$, $1955 \pm 285$ and $1394 \pm 343$  for solution II, respectively.
  The mass and width for the resonance around 1.47 GeV/${c^2}$ are determined to be $1477 \pm 7$~MeV/c$^2$ and $118 \pm 22$~MeV, respectively, which are consistent with those of the $\eta(1475)$ taken from PDG~\cite{pdg}.
 For the resonance around 1.83 GeV/${c^2}$, the obtained mass and width are $1839 \pm 26$~MeV/c$^2$ and $175 \pm 57$~MeV, respectively, which are consistent with the measurements of the $X(1835)$~\cite{X1835III,baihe}.

\begin{figure}[!htbp]
  \centering
  \includegraphics[width=0.42\textwidth]{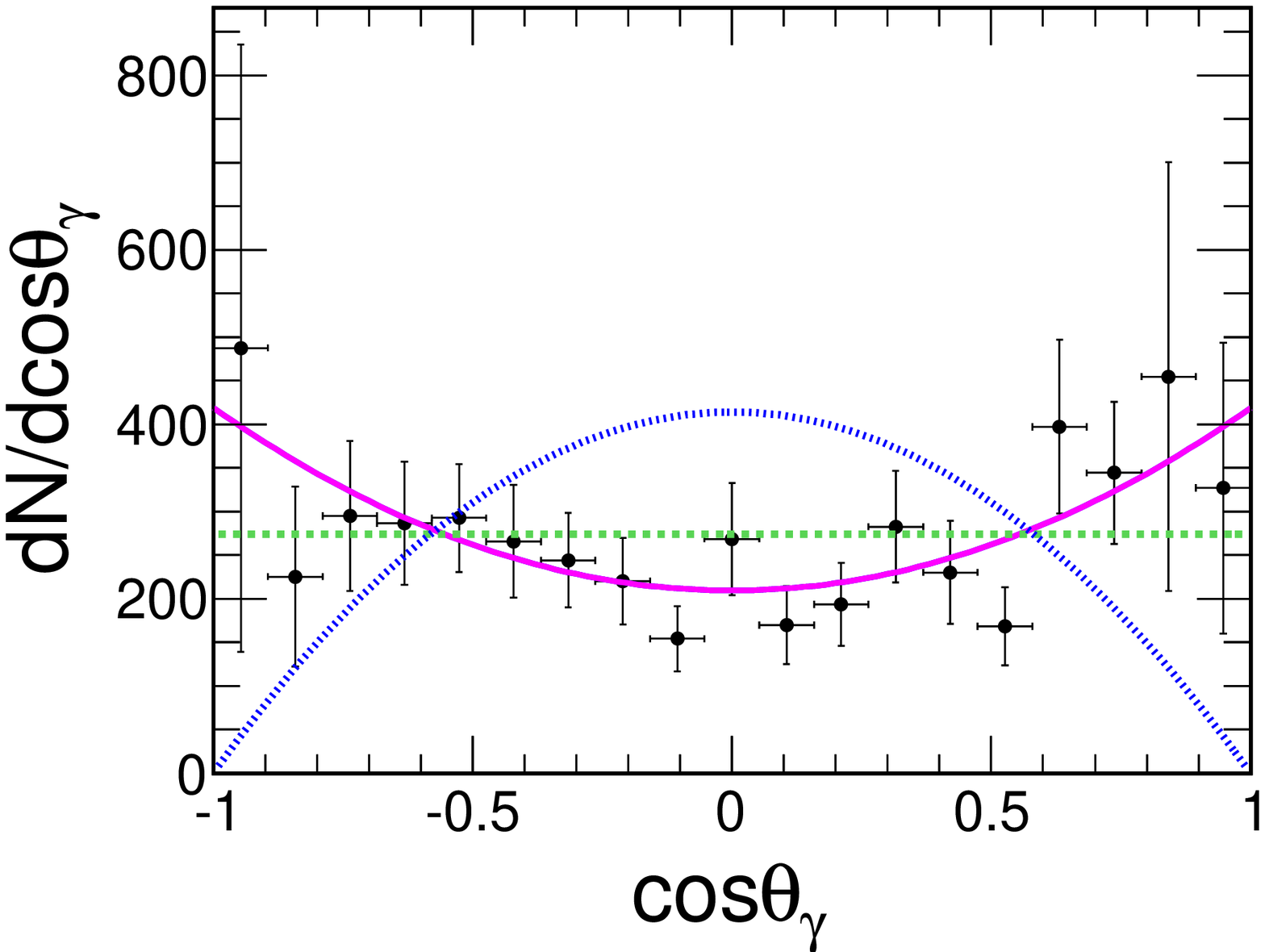}
   \put(-55,130){\textbf{\large{(a)}}}

  \includegraphics[width=0.42\textwidth]{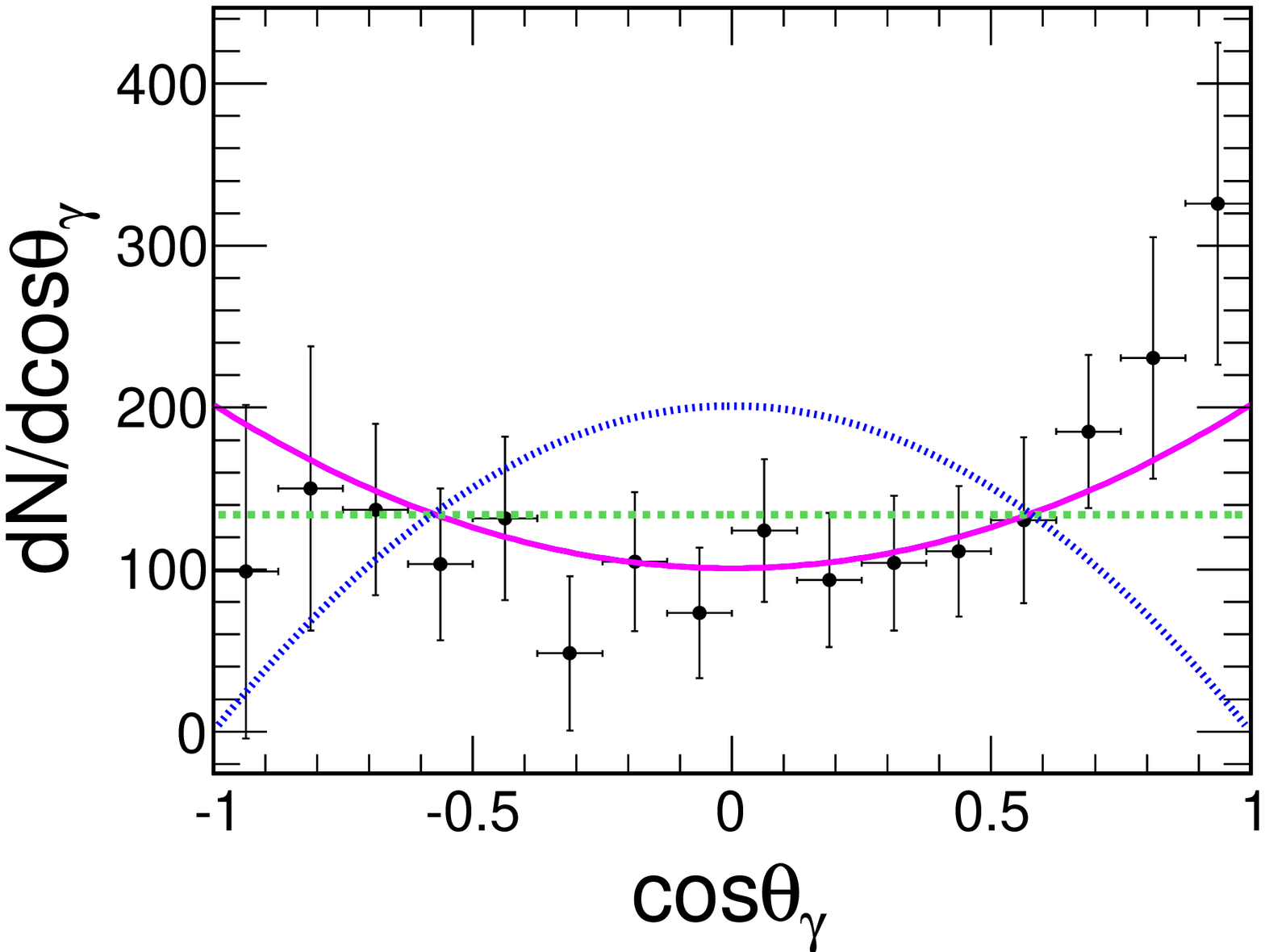}
  \put(-55,130){\textbf{\large{(b)}}}
  \caption{
   Fits to the efficiency-corrected $\cos\theta_{\gamma}$ distributions for (a) $1.4 < M(\gamma K+K-) < 1.6$~GeV/c$^{2}$ and (b) $1.75 < M(\gamma K+K-) < 1.9$~GeV/c$^{2}$. The dots with error bars represent data. The solid (pink), dashed (green) and dotted (blue) lines correspond to the hypotheses $\alpha=1$, 0 and $-1$, respectively.
  }
  \label{fig:angle}
\end{figure}

The polar angle distribution of the radiative photon in the $J/\psi$ rest system, $\cos\theta_{\gamma}$, is used to investigate the spin-parity $J^{PC}$ of the two new observed resonances.
The full $\cos\theta_{\gamma}$ range of [-1, 1] is divided into 19 and 16 bins for the candidate events within $1.4 < M(\gamma K^+ K^-) < 1.6$~GeV/$c^2$ and $1.75 < M(\gamma K^+ K^-) < 1.90$~GeV/$c^2$, respectively.
The signal yield in each $\cos\theta_{\gamma}$ bin is determined by a fit to the $M(K^+ K^-)$ spectrum as described above.
 The obtained $\cos\theta_\gamma$ distributions corrected for detection efficiency are shown in Fig.~\ref{fig:angle}.
For $J/\psi$ radiative decays to a pseudo-scalar meson, $\cos\theta_{\gamma}$ is distributed according to $(1+\alpha \cdot \cos^2\theta_{\gamma})$ with $\alpha = 1$.
Three least-$\chi^2$ fits are carried out on the $\cos \theta_\gamma$ distributions under the assumptions of $\alpha=-1, 0$, and $1$, respectively. As shown in Fig.~\ref{fig:angle}, the resulting $\chi^{2}/d.o.f$  for the resonance around 1.47 GeV/${c^2}$ are 152.0/18, 32.5/18, and 13.8/18 for $\alpha=-1, 0$, and $1$, respectively, which favor $\alpha = 1$ and a $J^{PC} = 0^{-+}$ assignment for this structure corresponding to $\eta(1475)$.
For the resonance around 1.83 GeV/${c^2}$, the resulting $\chi^2/d.o.f.$  are 55.8/15, 15.1/15, and 7.2/15 for $\alpha = -1, 0$, and $1$, respectively,  which favors $\alpha = 1$ and an assignment of $J^{PC} = 0^{-+}$ for the $X(1835)$ assumption.

Alternative fits are performed that include an additional $f_1(1420)$ or $\eta(1760)$ with mass and width fixed to the PDG values~\cite{pdg}.  They result in a statistical significance of less than $1.0\sigma$  for $f_1(1420)$ and $\eta(1760)$, respectively.
The statistical significance of the mass difference for the resonance around 1.47 GeV/${c^2}$  between the fit result and the world average value of the $\eta(1475)$~\cite{pdg} is calculated as $\chi^2_{\rm fix} - \chi^2_{\rm free}$ = $0.01$ corresponding to less than $0.1 \sigma$.
Here, $\chi^2_{\rm fix}$ and $\chi^2_{\rm free}$ are the chi-squared values of the fits with the mass fixed at the world average value of the $\eta(1475)$ and left free, respectively.
The statistical significance of the mass difference between the fit result and the world average value of the $\eta(1405)$~\cite{pdg} is $5.8 \sigma$.
The statistical significances of mass difference for the resonance around 1.83 GeV/${c^2}$  between the fit result and those in Refs.~\cite{X1835I, baihe} are both less than $1.0 \sigma$.

The branching fractions of $J/\psi \rightarrow \gamma X\rightarrow \gamma \gamma\phi$ are calculated as
\begin{equation}\label{bf}
   B(J/\psi \rightarrow \gamma X \rightarrow \gamma \gamma \phi)=\frac{N_{\rm obs}}{N_{J/\psi} \varepsilon B(\phi \rightarrow K^+K^-)},
\end{equation}
where $X$ is $\eta(1475)$ or $X(1835)$,
$N_{\rm obs}$ is the number of observed signal events determined from the fit to the $M(\gamma \phi)$ spectrum,
$N_{J/\psi}$ is the total number of $J/\psi$ events,
and $\varepsilon$ is the MC-determined detection efficiency which take into account the angular distribution.
$B(\phi \rightarrow K^+K^-)$ is the branching fraction of $\phi \rightarrow K^+K^-$ quoted from the PDG~\cite{pdg}.

The systematic uncertainties associate with the fit procedure arise from the fit range, signal shape and the non-resonant background contribution.
The uncertainty from the $\phi$ signal extraction is estimated by changing the $\phi$ fit regions in each $M(\gamma K^+ K^-)$ bins. The difference in the $\gamma\phi$ distributions is considered to be the systematic uncertainty.
In the nominal fit, the shapes of the $\eta(1475)$ and $X(1835)$ are described by Eq.~(\ref{BW}).
To estimate the uncertainties associated with the signal shape, we perform an alternate fit by replacing the signal shapes with $s$-dependent Breit-Wigner functions.
To estimate the uncertainties associated with the constraining, another fit without the constraing is performed, the difference between the two fits is considered to be the systematic uncertainty.
The bin size is changed from 35.00 to 33.75 and 36.35 MeV/$c^2$ and the maximum difference between the signal yields and the nominal values is taken as the systematic uncertainty.
To estimate the uncertainties associated with the $\phi$ background, the directly double radiative decay $J/\psi \rightarrow \gamma \gamma \phi$ is considered with MC simulated shape.

The systematic uncertainties on the branching fraction measurements are also subject to the uncertainties in the total number of $J/\psi$~\cite{njpsi} events,
the relevant branching fraction $B(\phi \rightarrow K^+ K^-)$ from the PDG~\cite{pdg},
kaon tracking,
kaon PID,
photon detection,
the kinematic fit, and the vetoes of $\pi^0$, $\eta$, and $\eta^{\prime}$.
The systematic uncertainties associated with the 5C kinematic fit are studied with the track helix parameter correction method, as described in Ref.~\cite{5csys}.
To estimate the uncertainties associated with the vetoes of $\pi^0$, $\eta$, and $\eta^{\prime}$, the gaussian functions are used to smear the $\phi \pi^0$, $\phi \eta$ and $\phi \eta^{\prime}$ MC simulated shapes to get a better consistent with data. The signals are smeared with the same parameters, and the difference between the smeared and unsmeared efficiencies are considered to be the systematic uncertainties.

Assuming all sources to be independent, the total systematic uncertainties on the product branching fractions of the $\eta(1475)$ and $X(1835)$ are determined by combining all the individual ones in quadrature.
The total systematic uncertainty on the product branching fraction of the $\eta(1475)$  is determined to be 12.9\% and 14.9\% for solution I and solution II, respectively. And it is determined to be 14.2\% and 16.8\% for the two solutions of $X(1835)$.
The systematic uncertainties on the mass and width of the $\eta(1475)$ and $X(1835)$ are estimated with a similar method.

Table~\ref{tab:solution1} lists the measured results.
The first uncertainties are statistical, and the second are systematic.
Since both combinations of $\gamma \phi$ are considered for each event without accounting for the associated statistical correlations, the uncertainties may be overestimated.
Although the significance of $f_1(1285)\to\gamma\phi$ is less than $5 \sigma$, the systematic uncertainty on its branching fraction is also estimated, and the result is shown in Table~\ref{tab:solution1}.

\begin{table}[!htbp]
\begin{small}
  \centering
  \caption{\label{tab:solution1} Mass, width, and $B(J/\psi\rightarrow \gamma X
    \rightarrow \gamma \gamma \phi)$ of each component in the
    two solutions (I) and (II). The first uncertainties are statistical and the second ones are
    systematic.}
  \linespread{1.5}
    \resizebox{0.48\textwidth}{!}{%
      \begin{tabular}{ccccc}
        \hline
        \hline
         Solution       &Resonance     &$m_R$\,(MeV/$c^{2}$)         &$\Gamma$\,(MeV)                      &$B$\,($10^{-6}$) \\
         \hline

                      &$\eta(1475)$ &~~ $1477\pm7 \pm13$   &~~$118\pm22\pm17$                       &~$7.03\pm0.92\pm0.91$                 \\
                      &$X(1835)$    &~~ $1839\pm26\pm26$   &~~$175\pm57\pm25$                       &~$1.77\pm0.35\pm0.25$               \\
        \hline

                      &$\eta(1475)$ &~~ $1477\pm7 \pm13$   &~~$118\pm22\pm17$                       &$10.36\pm1.51\pm1.54$                 \\
                      &$X(1835)$    &~~ $1839\pm26\pm26$   &~~$175\pm57\pm25$                       &~$8.09\pm1.99\pm1.36$               \\
        \hline
        \hline
      \end{tabular}
    }%
\end{small}
\end{table}

In summary, based on a sample of $1.31 \times 10^9$ $J/\psi$ events collected with the BESIII detector, we perform an analysis of the decay $J/\psi \rightarrow \gamma \gamma \phi$.
Two structures around 1.47 and 1.85 GeV/c$^2$ are observed in the $\gamma \phi$ invariant mass. A  fit on the $\gamma \phi$ invariant mass yields the resonant parameters and the decay branching fraction for the new observed structures as summarized in Table~\ref{tab:solution1}, and have statistical systematics of $13.5\sigma$ and $6.3\sigma$ for the structures around 1.47 and 1.85 GeV/c$^2$, respectively.
A fit on the polar angle distribution of the radiative photon favor $J^{PC} = 0^{-+}$ assignment for the two resonances. The obtained resonant parameters and $J^{PC}$ supports the two new observed resonances are $\eta(1475)$ and $X(1835)$, respectively, and this is for the first time we observed $\eta(1475)$ and $X(1835)$ decaying into $\gamma\phi$ final states.

 The partial width ratio of ($\Gamma(\eta(1405/1475) \rightarrow\gamma\rho)$ : $\Gamma(\eta(1405/1475) \rightarrow\gamma\phi)$) is calculated to be $(11.10\pm3.50)$ : 1 for the case of destructive interference and $(7.53\pm2.49)$ : 1 for constructive interference, where the branching fraction of $J/\psi \rightarrow \gamma \eta(1405/1475) \to \gamma \gamma \rho$ is taken from the BES measurement~\cite{eta1405obs}.
 The ratio is slightly larger than the prediction of 3.8 : 1 in Ref.~\cite{onestat2} for the case of a single pseudo-scalar state.
On the other hand, if the $\eta(1405)$ and the $\eta(1475)$ are different states, the observation of the $\eta(1475)$ decaying into $\gamma \phi$ final state suggests that the $\eta(1475)$ contains a sizable $s\bar{s}$ component and, if so, should be the radial excitation of the $\eta^{\prime}$~\cite{fluxtubemodel}.
The observation of the $X(1835)$ decaying into $\gamma \phi$ final state indicates that this resonance also contains a sizable $s\bar{s}$ component.
It seems therefore unlikely to be a pure $N\bar{N}$ bound state.

We extend our special thanks to J. J. Wu of the Special Research Centre for the Subatomic Structure of Matter (CSSM) for many helpful discussions.
The BESIII collaboration thanks the staff of BEPCII and the IHEP computing center for their strong support. This work is supported in part by National Key Basic Research Program of China under Contract No. 2015CB856700; National Natural Science Foundation of China (NSFC) under Contracts Nos. 11235011, 11322544, 11335008, 11425524, 11675183, 11175188, 11735014; the Chinese Academy of Sciences (CAS) Large-Scale Scientific Facility Program; the CAS Center for Excellence in Particle Physics (CCEPP); the Collaborative Innovation Center for Particles and Interactions (CICPI); Joint Large-Scale Scientific Facility Funds of the NSFC and CAS under Contracts Nos. U1232201, U1332201; CAS under Contracts Nos. KJCX2-YW-N29, KJCX2-YW-N45; 100 Talents Program of CAS; National 1000 Talents Program of China; INPAC and Shanghai Key Laboratory for Particle Physics and Cosmology; German Research Foundation DFG under Contracts Nos. Collaborative Research Center CRC 1044, FOR 2359; Istituto Nazionale di Fisica Nucleare, Italy; Joint Large-Scale Scientific Facility Funds of the NSFC and CAS	under Contract No. U1532257; Joint Large-Scale Scientific Facility Funds of the NSFC and CAS under Contract No. U1532258; Koninklijke Nederlandse Akademie van Wetenschappen (KNAW) under Contract No. 530-4CDP03; Ministry of Development of Turkey under Contract No. DPT2006K-120470; National Science and Technology fund; NSFC under Contract No. 11275266; The Swedish Resarch Council; U. S. Department of Energy under Contracts Nos. DE-FG02-05ER41374, DE-SC-0010118, DE-SC-0010504, DE-SC-0012069; U.S. National Science Foundation; University of Groningen (RuG) and the Helmholtzzentrum fuer Schwerionenforschung GmbH (GSI), Darmstadt; WCU Program of National Research Foundation of Korea under Contract No. R32-2008-000-10155-0.

\par

\end{document}